# Highly conductive PdCoO$_2$ ultrathin films for transparent electrodes

T. Harada,[1,a)] K. Fujiwara,[1] and A. Tsukazaki[1,2]

[1] *Institute for Materials Research, Tohoku University, Sendai 980-8577, Japan*

[2] *Center for Spintronics Research Network (CSRN), Tohoku University, Sendai 980-8577, Japan*

[a] Author to whom correspondence should be addressed: t.harada@imr.tohoku.ac.jp

**Abstract**

We report on the successful synthesis of highly conductive $PdCoO_2$ ultrathin films on $Al_2O_3$ (0001) by pulsed laser deposition. The thin films grow along the *c*-axis of the layered delafossite structure of $PdCoO_2$, corresponding to the alternating stacking of conductive Pd layers and $CoO_2$ octahedra. The thickness-dependent transport measurement reveals that each Pd layer has a homogeneous sheet conductance as high as 5.5 mS in the samples thicker than the critical thickness of 2.1 nm. Even at the critical thickness, high conductivity exceeding $10^4$ $Scm^{-1}$ is achieved. Optical transmittance spectra exhibit high optical transparency of $PdCoO_2$ thin films particularly in the near-infrared region. The concomitant high values of electrical conductivity and optical transmittance make $PdCoO_2$ ultrathin films as promising transparent electrodes for triangular-lattice-based materials.

A palladium-based delafossite metal, $PdCoO_2$, is one of the most conductive oxides characterized by long mean free path,[1,2] which gives rise to hydrodynamic electron flow in ultrapure single crystals.[3] The crystal structure of $PdCoO_2$, shown in Fig. 1(a), is a natural superlattice of highly conductive two-dimensional sheets of Pd atoms and triangular lattices of $CoO_2$ octahedra.[4] The Pd ions are weakly bound with oxygens to form O-Pd-O chains. This unique anisotropic crystal structure hosts remarkably high electrical conductivity with the bulk in-plane resistivity ~ 2.6 μΩcm at 300 K, which is even lower than that of Pd metal and comparable to that of Au.[5] Transport and optical measurements on bulk single crystals[5,6] as well as theoretical studies[7-10] have shown that the two-dimensional $Pd^{1+}$ layers are responsible for the high in-plane conductivity of $PdCoO_2$. Because of its high electrical conductivity, $PdCoO_2$ will be applicable to transparent electrodes when stabilized in an ultrathin film form to allow light to pass through, raising interests in thin-film growth of $PdCoO_2$. However, research of $PdCoO_2$ has been limited to bulk crystals[3-6, 11-17] and thick films prepared by post-annealing of amorphous precursors.[18] To date, ultrathin film growth of $PdCoO_2$ remains unexplored in literature.

Here, we demonstrate the thin-film growth of highly conductive $PdCoO_2$ on $Al_2O_3$ (0001) substrates. Figure 1(a) shows the crystal structure of $PdCoO_2$, consisting of three sets of the Pd and $CoO_2$ layers stacked alternately along the $c$-axis. The top views of the Pd and $CoO_2$ layers are illustrated in Figs. 1(b) and (c). The Pd and $CoO_2$ layers form triangular lattices with the in-plane lattice constants of $a = b = 2.83$ Å (Ref. 4). The oxygen atoms in $PdCoO_2$, shown as red spheres in Fig. 1(c), form a triangular lattice with the inter-oxygen distance of 2.83 Å that equals to the in-plane lattice constants. This inter-oxygen distance in $PdCoO_2$ is close to that in $Al_2O_3$ (2.75 Å on average), suggesting the $Al_2O_3$ (0001) an appropriate substrate for growth of $PdCoO_2$ thin films. A consideration of the epitaxial relationship between $PdCoO_2$ and $Al_2O_3$ suggests two possible configurations, $PdCoO_2$ (A) and $PdCoO_2$ (B), as shown in Fig. 1(d): the orientation of oxygen triangles on the $PdCoO_2$ surface is in-plane upward in the domain A and is rotated by 180° in the domain B. Although the crystal structure is different between delafossite $PdCoO_2$ and corundum $Al_2O_3$, the close similarities in the triangular lattice motif of surface oxygens enable us to synthesize highly crystalline $PdCoO_2$ thin films.

We have prepared $PdCoO_2$ thin films by pulsed-laser deposition (PLD). Commercially available $Al_2O_3$ (0001) substrates (SHINKOSHA CO., LTD) were ultrasonically cleaned by acetone and ethanol, followed by annealing in air at 900 °C for 12 hours to obtain an atomically flat surface with step-and-terrace structures. The annealed substrate was loaded into a vacuum chamber and preheated at $T = 700$ °C for 10 min under an oxygen partial pressure of $P_{O2} = 100$ mTorr. A stoichiometric $PdCoO_2$ target (KOJUNDO CHEMICAL LABORATORY Co., LTD) and a Pd-PdO mixed-phase target, prepared by sintering pelletized PdO powder at 1000 °C for 24 h in air, were ablated alternately by KrF excimer laser with the laser fluence of 2 J/cm$^2$ under the growth condition of $T = 700$ °C and $P_{O2} = 100$ mTorr. We repeated an ablation sequence of 140 pulses at the laser repetition rate of 5 Hz for $PdCoO_2$ and 300 pulses at 15 Hz for Pd-PdO targets. This single cycle resulted in 0.2 nm deposition on average. We repeated the cycle to fabricate the thin films with desired thickness. The samples were cooled down to room temperature in roughly 10 min immediately after the growth. The cation composition in the films was evaluated by inductively coupled plasma atomic emission spectroscopy (ICP-AES) and energy-dispersive X-ray spectroscopy (EDX) using the calibration line determined by the ICP-AES of a thick ($d \sim 200$ nm) sample. We note that thin films prepared with only a $PdCoO_2$ target suffer from significant Pd deficiencies with typical composition ratios of Pd/Co ~ 0.6. Employing the alternate deposition with the Pd-PdO target was found to effectively improve the stoichiometry as Pd/Co = 0.9 ± 0.1. The surface morphology was measured by atomic force microscopy (AFM). Crystal structures of the $PdCoO_2$ thin films were characterized by X-ray diffraction (XRD). The thickness $d$ of the samples was determined by the thickness fringes observed around the $PdCoO_2$ (0006) diffraction peaks. High-resolution transmission electron microscope (HRTEM) images were collected with a JEOL EM-002B. For electrical transport measurement, indium electrodes were mechanically soldered on the samples. The temperature dependence of the sample resistance was measured by a four-terminal method using a Quantum Design, physical property measurement system (PPMS). The transmittance spectra were measured with a Shimadzu UV-3600 plus equipped with an integrating sphere MPC-3100.

The XRD patterns for the $PdCoO_2$ thin films with different thicknesses are shown in Figure 2(a). All the film peaks are assigned to $PdCoO_2$ (000$\underline{3n}$), indicating that $c$-axis oriented $PdCoO_2$ thin films are grown without any traces of impurity phases. The widths of the $PdCoO_2$ (000$\underline{3n}$) peaks in the $2\theta$-$\omega$ scans become larger with decreasing the $PdCoO_2$ thickness, as expected from the Laue function. In addition, clear interference thickness fringes appear around the main diffraction peaks for all the samples. We determined the film thickness denoted as the numbers in Fig. 2(a) using the periodicity of the fringes near $PdCoO_2$ (0006) peaks. To identify the epitaxial relationship of $PdCoO_2$ and $Al_2O_3$, we measured the XRD $\phi$-scans around the $PdCoO_2$ (01$\bar{1}$2) and $Al_2O_3$ (01$\bar{1}$2) diffraction peaks. As shown in Fig. 2(b), the $\phi$-scan diffraction patterns have a 6-fold symmetry for $PdCoO_2$ and a 3-fold symmetry for $Al_2O_3$. The peak positions in the $\phi$-scans of the $PdCoO_2$ (01$\bar{1}$2) are shifted by $\Delta\phi = 30°$ from that of $Al_2O_3$ (01$\bar{1}$2), indicating the epitaxial relationship with the same oxygen triangular configuration in Fig. 1(d), where the lattice unit of $PdCoO_2$ and $Al_2O_3$ are 30-degree rotated from each other. The $PdCoO_2$ thin films showed the 6-fold $\phi$-scan diffraction pattern, despite the 3-fold symmetric crystal structure of delafossite $PdCoO_2$. It is thus reasonable to consider that the $PdCoO_2$ thin film has two-domains with the in-plane orientation different by $\Delta\phi = 180°$. This is supported by the AFM image of the $PdCoO_2$ surface shown in Fig. 2(c), which detects triangular shapes aligned to two directions with one of the base parallel to the [1$\bar{1}$00]$_{Al2O3}$ direction. These triangular domains are consistent with the two kinds of crystalline orientations found in the XRD $\phi$-scan and domains A and B in Fig. 1(d).

The lattice periodicity in the $PdCoO_2$ thin film was resolved by HRTEM. As shown in Fig. 2(d), the HRTEM image of a $PdCoO_2$ thin film shows periodic bright lines stacked along the $PdCoO_2$ [0001] direction, corresponding to the layered crystal structure of the $PdCoO_2$ thin film. The period of the bright lines is about 0.59 nm, which agrees well with the 1/3 unit cell height, *i.e.*, a single stack of Pd layer and $CoO_2$ octahedron. This periodicity is clearly observed in the diffraction pattern of $PdCoO_2/Al_2O_3$ shown in the inset of Fig. 2(d). Clear $PdCoO_2$ (000$\underline{3n}$) bright spots, indicated by white arrows, are observed together with those for the $Al_2O_3$ substrate indicated by gray arrows. We note that the $Al_2O_3$ substrate showed normally forbidden (0003) and (000$\bar{3}$) spots, which might be due to strained crystal structure with off-stoichiometric oxygen composition[19] caused by annealing in growth process. The periodic HRTEM image evidences that the atomic layers in $PdCoO_2$ are regularly ordered along the $PdCoO_2$ [0001] direction.

The thickness dependence of the room-temperature sheet conductance ($1/R_s^{300K}$) is plotted in Fig. 3(a). The $1/R_s^{300K}$ linearly increases above the thickness larger than $d = 2$ nm, indicating that each $PdCoO_2$ layer possesses the homogeneous sheet conductance. The room-temperature sheet conductance per Pd sheet, deduced from the slope of the fitting line in Fig. 3(a), is as high as 5.5 mS, which is comparable to that of the doped graphene (~ 8 mS).[20] The existence of a roughly 1-unit-cell dead layer implies that the initial growth unit seems a whole unit cell of $PdCoO_2$ composed of three sequences of the Pd sheet/$CoO_2$ layers, rather than the 1/3 unit cell, *i.e.*, a single sequence of the Pd sheet / $CoO_2$ stack. Formation of such a whole single unit cell might be a key requirement to stabilizing the structure and producing high sheet conductance. After the formation of the 1-unit-cell initial layer, growth unit seems to become 1/3 unit cell as evidenced by steps with the 1/3-unit-cell height in the cross-sectional height profile in Fig. 2(c). We note that a 1-unit-cell-thick dead layer should exist also in the thicker films ($d > 2$ nm), as indicated by the linear $1/R_s$ vs $d$ dependence shifted to the positive direction in $d$ axis. Extrinsic scattering origins such as surface roughness and interfacial effects could be responsible for the persisting dead layer.

The temperature dependence of the sheet resistance ($R_s$-$T$ curve) for $d = 2.1 – 8.8$ nm is displayed in Fig. 3(b). All the samples show monotonic decrease of $R_s$ upon cooling down to $T \sim 40$ K with small upturns at low temperature. In comparison with polycrystalline and single-crystalline bulk studies, residual resistance ratio (RRR) of about 2 in our films is closer to that in polycrystalline bulk of about 4,[21] rather than high purity single-crystal of 400.[14] The plausible origins for such upturns are either impurity in the thin films or carrier scattering at the domain boundaries. Further improvement of crystallinity in PdCoO$_2$ thin films will lead to increase $1/R_s^{300K}$ per Pd sheet toward the bulk corresponding value of ~23 mS,[1] by reducing the effect of impurities and the domain boundaries.[14]

The optical transmittance spectra of the PdCoO$_2$/Al$_2$O$_3$ in Fig. 4 represent the specific features of the band structure and high optical transparency. The overall spectral shape is identical for all the thicknesses, representing high transparency at near-infrared region at around 1 eV. The transmittance spectra have three characteristic dips around 1.2, 2.8, and 5.3 eV (red arrows), positions of which are independent of the PdCoO$_2$ film thickness. As discussed for noble metals[22] and correlated oxides,[23] inter-band electronic transition between flat filled band at some different $k$-points and empty band just above the Fermi surface might be the origin of the observed dip structures. The dip positions in the transmittance spectra in Fig. 4 are consistent with the flat band positions in $k$-scape calculated by density functional theory, which is located lower than the Fermi level by $E_1$ ~ 0.8-1.5; $E_2$ ~ 2.3-3.0; $E_3$ ~ 5.6-5.8 eV.[10] The PdCoO$_2$ ultrathin films possess high optical transparency as shown in the photograph in the inset of Fig. 4. To evaluate the transmittance of the PdCoO$_2$ films $T_{film}$ for three energy regions (three color lines in Fig. 4), the optical loss in the Al$_2$O$_3$ is subtracted as $T_{film} = T_{film+sub}/T_{sub}$. The thickness dependence of the $T_{film}$ at the selected wavelengths in near-infrared (NIR), visible (Vis), and ultraviolet (UV) regions are plotted in the left inset of Fig. 4. The $T_{film}$ has an approximately linear dependence on the PdCoO$_2$ thin film $d$. In particular, at the energy of 0.8 eV in the NIR region, the $T_{film}$ is as high as 80 % even for the thick PdCoO$_2$ thin films ($d$ = 8.8 nm), demonstrating the superior optical factor as a NIR transparent conductor.

In summary, PdCoO$_2$ ultrathin films are successfully synthesized on Al$_2$O$_3$ (0001) substrates by PLD. High sheet conductance is obtained above $d$ = 2.1 nm, the thickness of which is close to the $c$-axis lattice unit. The PdCoO$_2$ thin films have two domains, depending on the configurations of triangular CoO$_2$ lattices on an Al$_2$O$_3$ (0001) surface. Ultrathin PdCoO$_2$ possesses high optical transparency, particularly in NIR region, while keeping low sheet resistance ~ 100 Ω. Further attempts to reduce the low-temperature residual resistance by suppressing impurities and domain boundaries will broaden the applicable research field of PdCoO$_2$ thin film, for example, mesoscopic systems and heterostructures.

**Acknowledgement**

The authors thank Dr. S. Kuboya and Prof. T. Matsuoka for the help in optical transmittance measurement, Mr. S. Ito for the HRTEM measurement, and Mr. F. Sakamoto for ICP-AES analysis. This work is a cooperative program (Proposal No. 16G0404) of the CRDAM-IMR, Tohoku University. This work is partly supported by a Grant-in-Aid for Specially Promoted Research (No. 25000003), a Grant-in-Aid for Scientific Research (A) (No. 15H02022) from the Japan Society for the Promotion of Science (JSPS), and Mayekawa Houonkai Foundation.

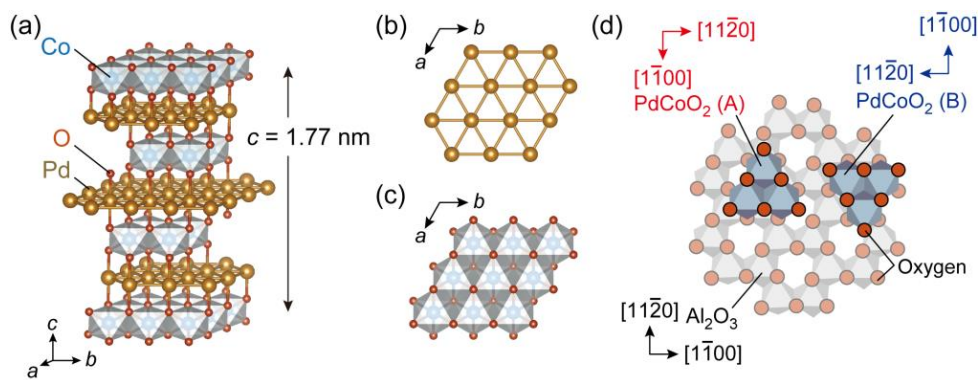

**FIG. 1.** (a) Crystal structure of PdCoO$_2$. The *c*-axis length is about 1.77 nm. In-plane lattice configurations of (b) the Pd layer and (c) the CoO$_2$ layers. (d) Schematic illustration of possible epitaxial relationships of PdCoO$_2$ on an Al$_2$O$_3$ (0001) substrate surface. The oppositely placed CoO$_2$ layers (domain A and B) are depicted.

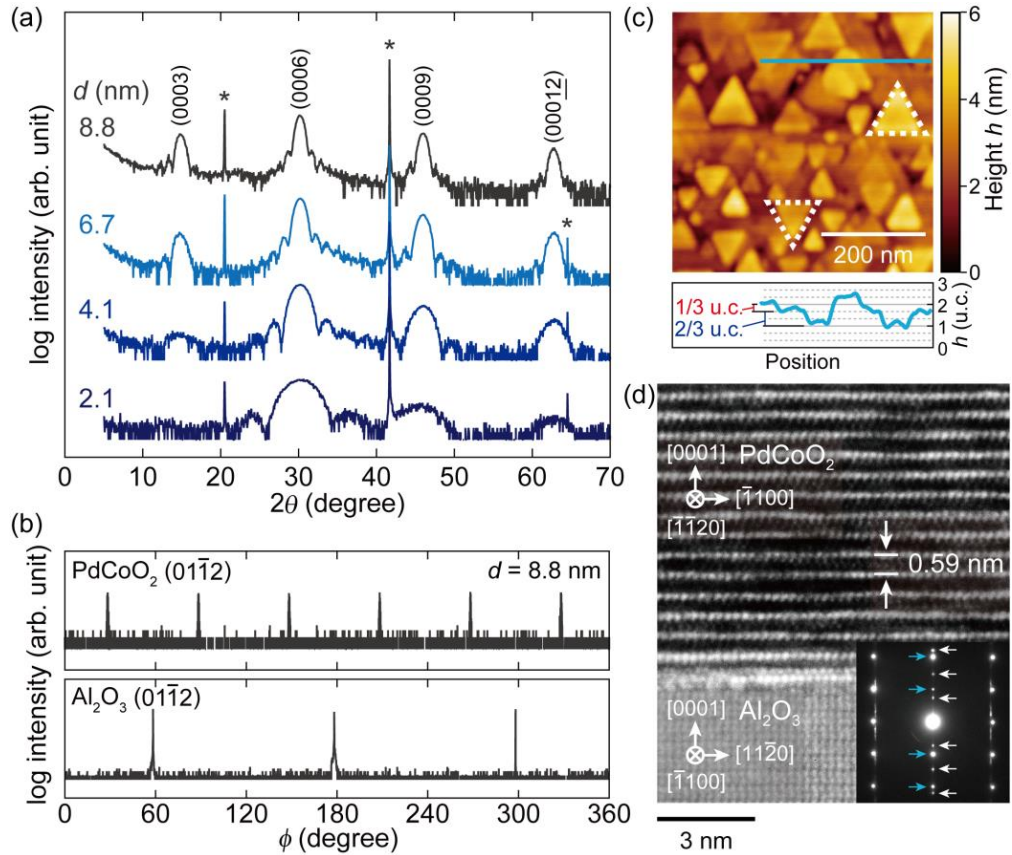

**FIG. 2.** (a) The thickness-dependent $2\theta$-$\omega$ XRD patterns of PdCoO$_2$ thin films grown on Al$_2$O$_3$ (0001). The substrate peaks are indicated by asterisks (*). The numbers in the figure are thickness $d$ (nm). (b) The XRD in-plane azimuthal scans for the PdCoO$_2$ (8.8 nm)/Al$_2$O$_3$ around the PdCoO$_2$ (01$\bar{1}$2) (top) and the Al$_2$O$_3$ (01$\bar{1}$2) diffraction peaks (bottom). (c) The AFM topographic image of the PdCoO$_2$ (6.7 nm)/Al$_2$O$_3$. Two triangular shapes corresponding to two domains are highlighted by the white dotted lines. The cross-sectional height profile along the light blue line is shown at the bottom. The steps with 1/3 and 2/3 unit-cell (u.c.) height are marked. The scale of the bottom axis for the height profile is same as the AFM image. (d) Cross-sectional HRTEM image of PdCoO$_2$ (10.7 nm)/Al$_2$O$_3$ in Al$_2$O$_3$ [1$\bar{1}$00] projection measured under the electron acceleration voltage of 200 kV. The inset shows the diffraction pattern measured over a wide region including the PdCoO$_2$ thin film and the Al$_2$O$_3$ substrate. The PdCoO$_2$ (000$\underline{3n}$) and Al$_2$O$_3$ (000$\underline{3n}$) reflections are indicated by white and blue arrows, respectively.

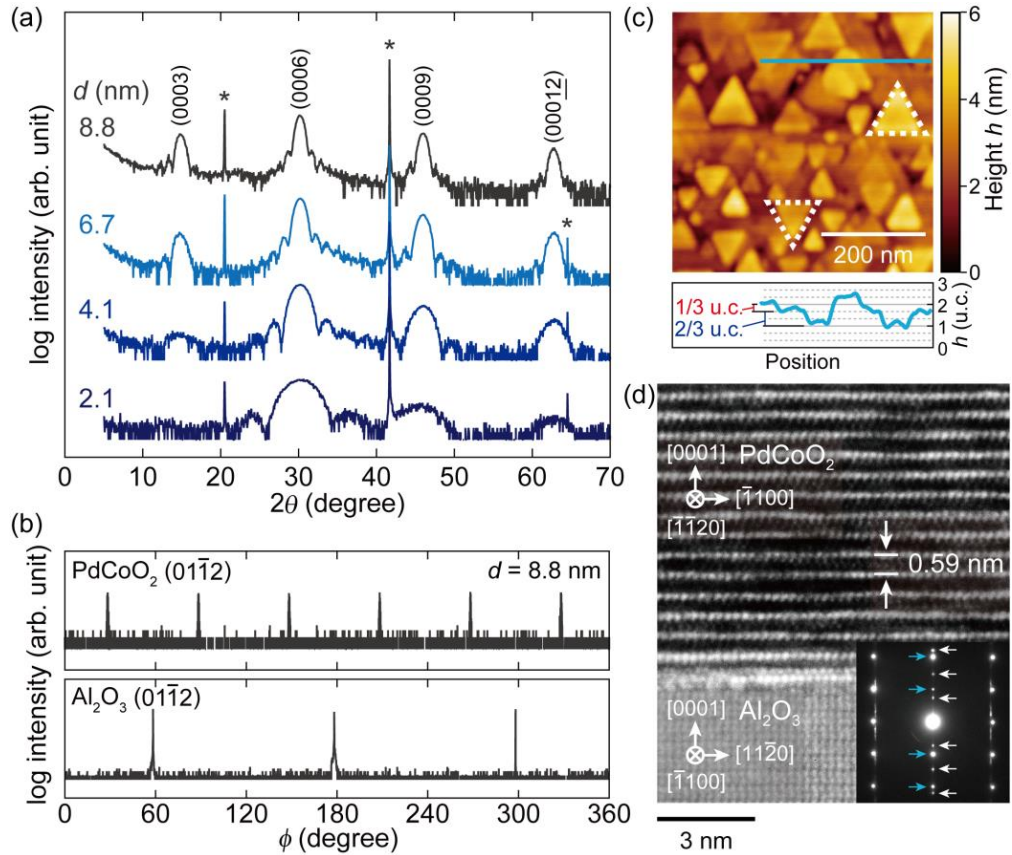

**FIG. 2.** (a) The thickness-dependent $2\theta$-$\omega$ XRD patterns of PdCoO$_2$ thin films grown on Al$_2$O$_3$ (0001). The substrate peaks are indicated by asterisks (*). The numbers in the figure are thickness $d$ (nm). (b) The XRD in-plane azimuthal scans for the PdCoO$_2$ (8.8 nm)/Al$_2$O$_3$ around the PdCoO$_2$ (01$\bar{1}$2) (top) and the Al$_2$O$_3$ (01$\bar{1}$2) diffraction peaks (bottom). (c) The AFM topographic image of the PdCoO$_2$ (6.7 nm)/Al$_2$O$_3$. Two triangular shapes corresponding to two domains are highlighted by the white dotted lines. The cross-sectional height profile along the light blue line is shown at the bottom. The steps with 1/3 and 2/3 unit-cell (u.c.) height are marked. The scale of the bottom axis for the height profile is same as the AFM image. (d) Cross-sectional HRTEM image of PdCoO$_2$ (10.7 nm)/Al$_2$O$_3$ in Al$_2$O$_3$ [1$\bar{1}$00] projection measured under the electron acceleration voltage of 200 kV. The inset shows the diffraction pattern measured over a wide region including the PdCoO$_2$ thin film and the Al$_2$O$_3$ substrate. The PdCoO$_2$ (000$\underline{3n}$) and Al$_2$O$_3$ (000$\underline{3n}$) reflections are indicated by white and blue arrows, respectively.

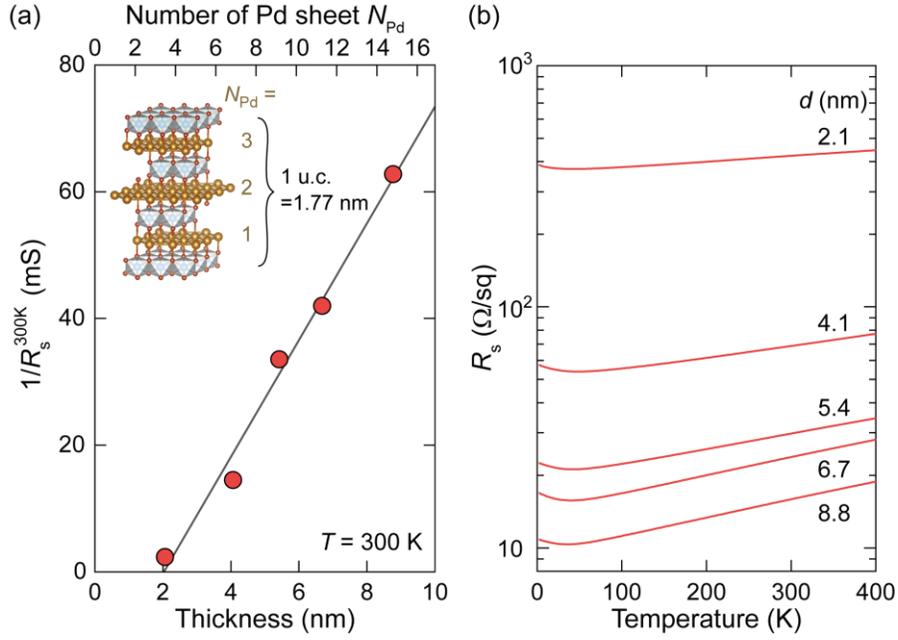

**FIG. 3.** (a) The thickness dependence of the room-temperature sheet conductance $1/R_s^{300K}$. The inset shows 1-u.c.-thick $PdCoO_2$, which contains 3 Pd sheets. (b) The $R_s$-$T$ curves for $PdCoO_2$ thin films with different thickness. The numbers in the panel correspond to the $PdCoO_2$ thickness $d$ (nm).

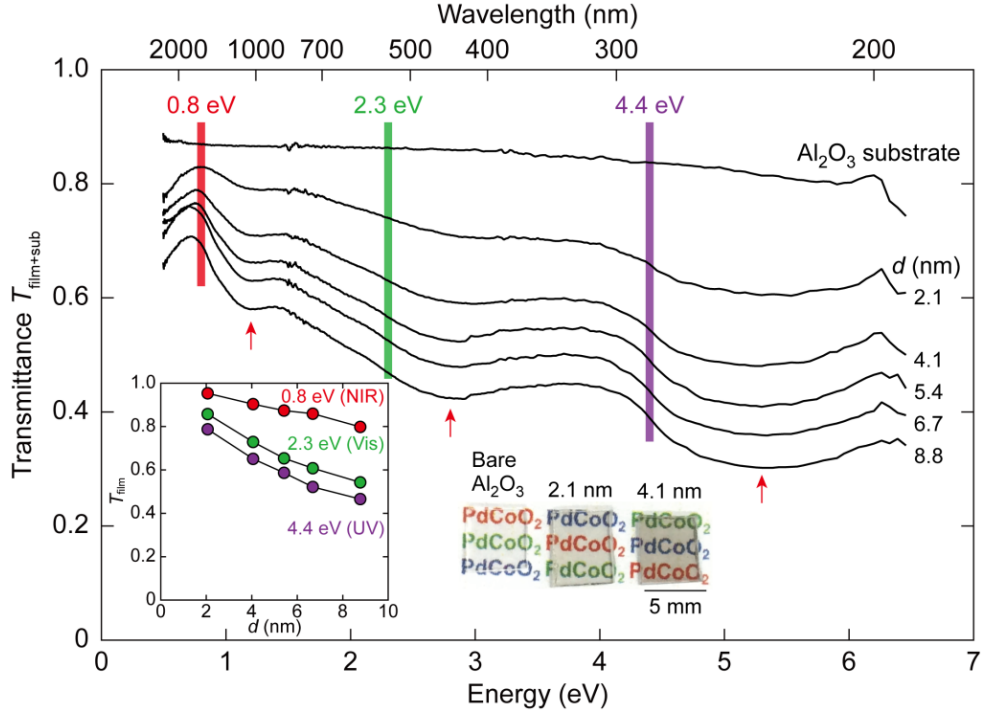

**FIG. 4.** Transmittance spectra of PdCoO$_2$ ($d$ nm)/Al$_2$O$_3$. The transmittance of the bare Al$_2$O$_3$ substrate without PdCoO$_2$ layers is also represented. The red arrows indicate characteristic dip structures. The left inset shows the thickness dependence of the transmittance of PdCoO$_2$ film ($T_{film} = T_{film+sub}/T_{sub}$) plotted for selected photon energies: 0.8 eV (NIR: red), 2.3 eV (Vis: green), and 4.4 eV (UV: purple). The corresponding photon energies are indicated by vertical lines in the main transmittance spectra. The right inset is the photograph of the bare Al$_2$O$_3$ substrate (left), PdCoO$_2$ (2.1 nm)/Al$_2$O$_3$ (middle), and PdCoO$_2$ (4.1 nm)/Al$_2$O$_3$ (right).